\documentclass[letterpaper, 10 pt, conference]{ieeeconf}  
\IEEEoverridecommandlockouts                              
\usepackage{fancyhdr}
\usepackage{graphics} 
\usepackage{epsfig} 
\usepackage{epstopdf}            
\usepackage{times} 
\usepackage{amsmath}   
\usepackage{amssymb}   
\usepackage{xcolor}
\usepackage{graphicx}
\usepackage[letterpaper, left=0.75in, right=0.75in, bottom=0.75in, top=0.75in]{geometry}
\usepackage{tikz}
\usepackage{booktabs}
\usepackage{flushend}
\usepackage{hyperref}

\newcommand{\volume}{{\ooalign{\hfil$V$\hfil\cr\kern0.08em--\hfil\cr}}}
\title{\LARGE \bf Design of a Ballistically-Launched Foldable Multirotor}

\author{Daniel Pastor$^{1}$, Jacob Izraelevitz$^{2}$, Paul Nadan$^{3}$, Amanda Bouman$^{1}$,  Joel Burdick$^{1}$, and Brett Kennedy$^{2}$
\thanks{$^{1}$Daniel Pastor, Amanda Bouman and Joel Burdick are with the Department of Mechanical and Civil Engineering,
        California Institute of Technology, Pasadena, CA 91125, USA,
        {\tt\small dpastorm@caltech.edu, abouman@caltech.edu, jwb@robotics.caltech.edu}}%
\thanks{$^{2}$Jacob Izraelevitz and Brett Kennedy are with the Jet Propulsion Laboratory, California Institute of Tehcnology, Pasadena, CA 91109 
        {\tt\small jacob.izraelevitz@jpl.nasa.gov, bkennedy@jpl.nasa.gov}}%
\thanks{$^{3}$Paul Nadan is with the Olin College, Needham, MA 02492, 
        {\tt\small paul.nadan@students.olin.edu}}%
\thanks{DISTRIBUTION STATEMENT A (Approved for Public Release, Distribution Unlimited)}%
}

\begin{document}
\vspace{0.25in}
\maketitle
\thispagestyle{fancy}
\thispagestyle{empty}
\pagestyle{empty}

\begin{abstract}
The operation of multirotors in crowded environments requires a highly reliable takeoff method, as failures during takeoff can damage more valuable assets nearby. The addition of a ballistic launch system imposes a deterministic path for the multirotor to prevent collisions with its environment, as well as increases the multirotor's range of operation and allows deployment from an unsteady platform. In addition, outfitting planetary rovers or entry vehicles with such deployable multirotors has the potential to greatly extend the data collection capabilities of a mission. A proof-of-concept multirotor aircraft has been developed, capable of transitioning from a ballistic launch configuration to a fully controllable flight configuration in midair after launch. The transition is accomplished via passive unfolding of the multirotor arms, triggered by a nichrome burn wire release mechanism. The design is 3D printable, launches from a three-inch diameter barrel, and has sufficient thrust to carry a significant payload. The system has been fabricated and field tested from a moving vehicle up to 50mph to successfully demonstrate the feasibility of the concept and experimentally validate the design's aerodynamic stability and deployment reliability.
\end{abstract}


\section*{Supplementary Material}
Videos of the experiments: \url{https://youtu.be/sQuKJfllyRM}

\section{Introduction} \label{sec:Background}

There is growing interest in developing ballistically launched small aircraft systems (sUASs), for applications in both emergency response and space exploration. Thus far, successful systems have been implemented for both fixed wing aircraft~\cite{raytheonCoyote:online}\cite{uvisionHero:online}\cite{leonardo:online}  and coaxial rotorcraft~ \cite{gnemmi2017conception}, and multirotor designs are starting to enter development~\cite{henderson2017towards}. The Streamlined Quick Unfolding Investigation Drone (SQUID) design detailed in this paper provides a multirotor implementation of the ballistically launched sUAS concept. One of the primary motivations for this project is to fulfill the need from emergency response and security teams to quickly deploy a multirotor from a moving vehicle in order to provide support and coverage.

\begin{figure}[t]
	\centering
	\includegraphics[width=0.32\linewidth, trim={850 210 880 400},clip]{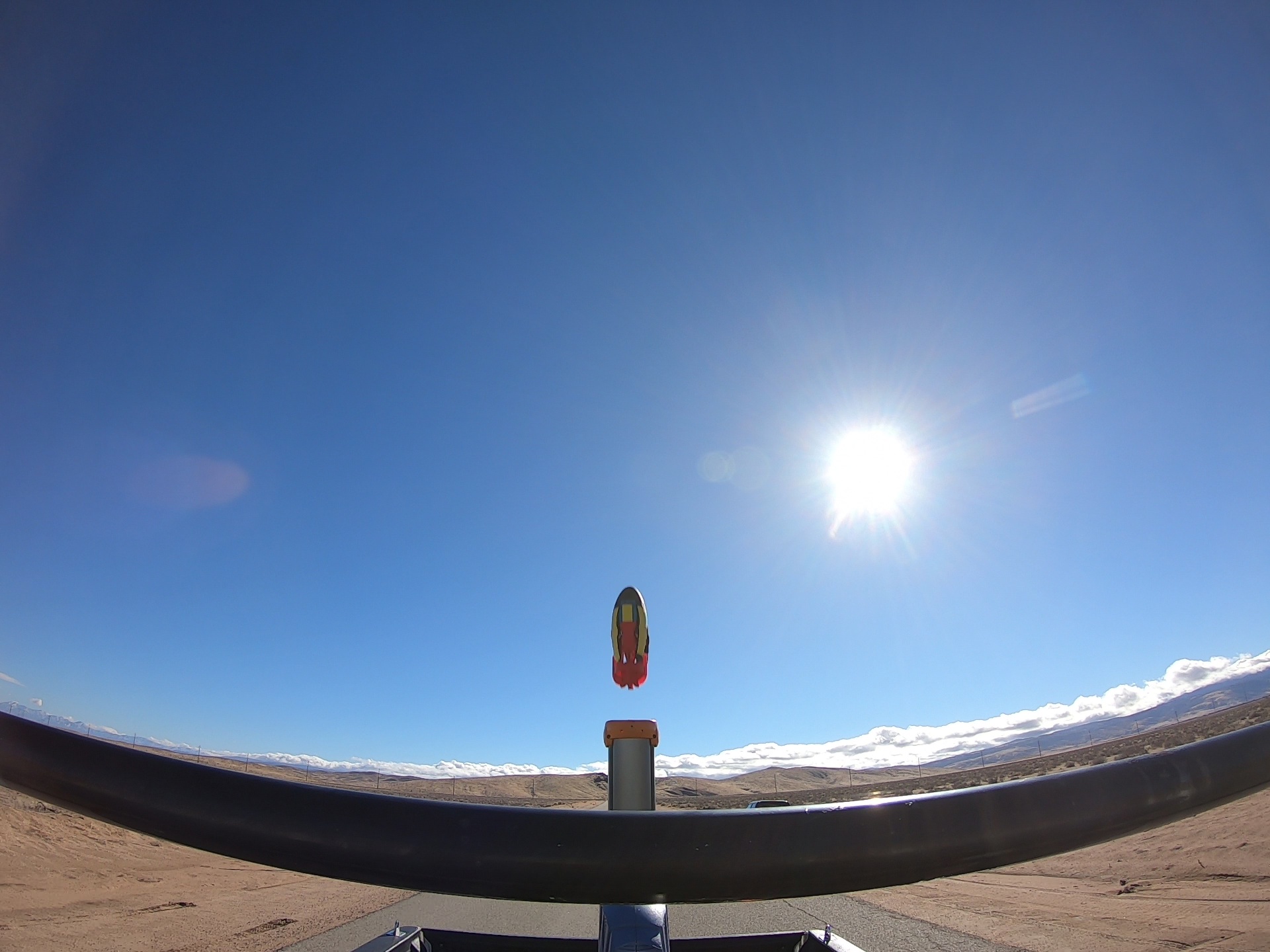}
	\includegraphics[width=0.32\linewidth, trim={850 210 880 400},clip]{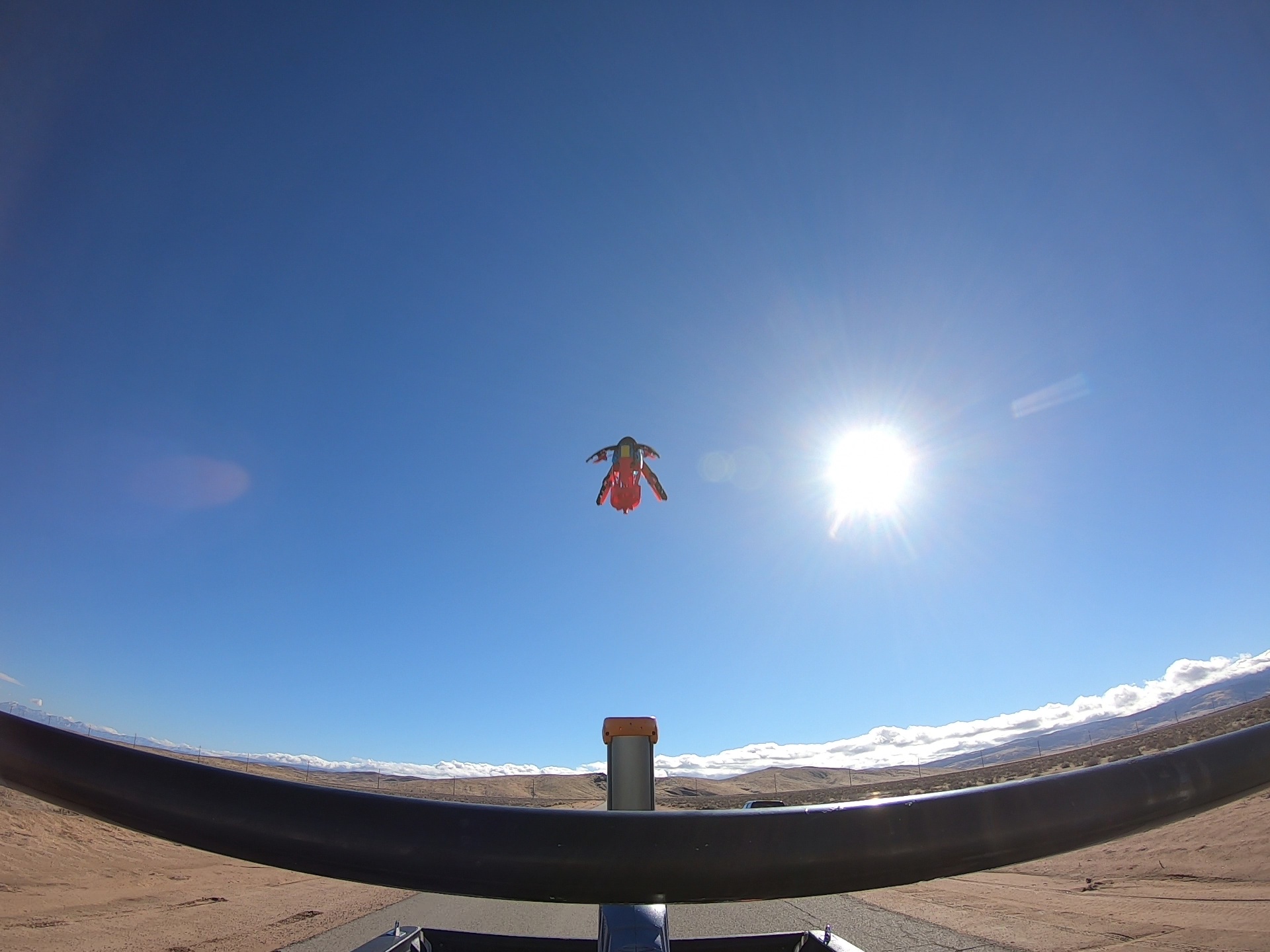}
	\includegraphics[width=0.32\linewidth, trim={850 210 880 400},clip]{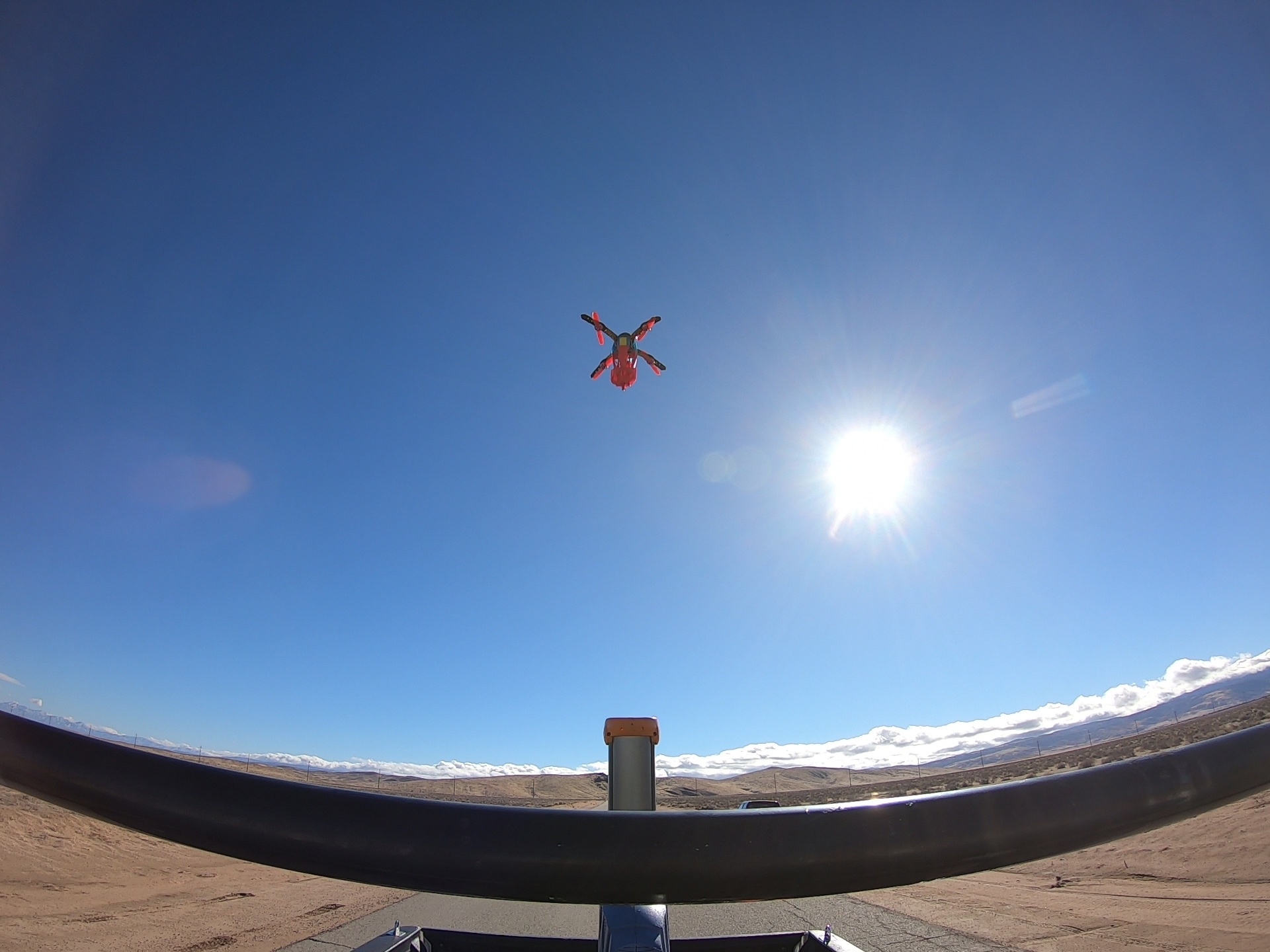}
	\caption{The SQUID prototype in both ballistic (left), deploying its arms (center) and multirotor (right) configurations during flight. Selected frames during a launch from a moving vehicle at 50mph (22 m/s). After deployment the drone hovered around the area.}
	\label{fig:car_video_frames}
\end{figure}

Takeoff is one of the most dangerous portions of a multirotor's flight, as it involves hazards to not only the multirotor but also other assets on the ground. A ballistic launch addresses this problem by creating a pre-determined path for the multirotor away from higher-value assets, even in the case of aircraft failure. A typical scenario would involve deployment from a windy roof, the bed of a truck, or a ship in waves. In these scenarios, the vehicle is stored for long periods of time and must quickly provide air support in the case of an unexpected event. Current drone designs are slow to deploy, require user intervention prior to takeoff, and cannot be deployed from a moving vehicle. Furthermore, traditional foldable designs require the user to unfold the arms, slowing the process and putting the user at risk. In the case of deployment from a moving vehicle, the drone also needs to be aerodynamically stable to avoid tumbling when exposed to sudden crosswinds. The design, development and testing of such a vehicle is the main contribution of this paper.

There are several drones in the market that can be folded to occupy a small volume, the most popular being the Mavic series from DJI~\cite{Mavic2th36:online} and Anafi from Parrot~\cite{DroneCam11:online}. Both multirotors rotate their arms horizontally to fold into a small volume. This illustrates a considerable interest in foldable multirotors. However, these designs cannot fit smoothly inside of a launch system, and the unfolding is manual and not automatic. The following drones can also fold into a form factor similar to a cylinder like SQUID: the \textit{Power Egg} from Power Vision folds into an egg shaped drone~\cite{PowerEgg6:online}, the drone from LeveTop folds into a small cylinder~\cite{LeveTopT73:online}, and Sprite from Ascent Aerosystems has a coaxial design and it can be packed into a cylinder shape~\cite{AscentAe83:online}. Other designs get inspiration from origami~\cite{origamiCargo}. To compare them to a ballistically-launched drone like SQUID, all past foldable designs would have to be redesigned to withstand the launch loads, the autopilot would have to be reconfigured, and a mechanism must be added to automatically deploy the arms and/or propellers. For fixed wing aircraft, there are several mature products for military applications, notably Coyote from Raytheon with two sets of wings~\cite{raytheonCoyote:online}, Hero series from UVision with its X-shaped wings\cite{uvisionHero:online}, and Horus from Leonard which can be launched from a tank\cite{leonardo:online}.


While the SQUID prototype, as outlined in this paper, has been designed for operation on Earth, the same concept is potentially adaptable to other planetary bodies, in particular Mars and Titan. The Mars helicopter, planned to deploy from the Mars 2020 rover, will provide a proof-of-concept for powered rotorcraft flight on the planet, despite the thin atmosphere~\cite{bib:mars}. A rotorcraft greatly expands the data collection range of a rover, and allows access to sites that a rover would find impassible. However, the current deployment method for the Mars Helicopter from the underbelly of the rover reduces ground clearance, resulting in stricter terrain constraints. Additionally, the rover must move a significant distance away from the helicopter drop site before the helicopter can safely take off. The addition of a ballistic, deterministic launch system for future rovers or entry vehicles would isolate small rotorcraft from the primary mission asset, as well as enable deployment at longer distances or over steep terrain features. Titan is another major candidate for rotorcraft flight. The Dragonfly mission proposal to the New Frontiers Program illustrates how rotorcraft can take advantage of the thick atmosphere and low gravity of Titan to fly to many different sites with the same vehicle~\cite{bib:titan}. A SQUID-type launch applied to Titan could be used for deployment of small daughter rotorcraft from landers, airships, or lake buoys, expanding the option space for Titan mission design.





This paper is organized as follows: Section~\ref{sec:Design} will cover the design of the vehicle, Section~\ref{sec:Operations} will describe the operations, Section~\ref{sec:Testing} will cover the main testing demonstrations, Section~\ref{sec:Scaling} will describe scaling arguments, and conclusions are presented in Section~\ref{sec:Conclusion}.
\section{Vehicle Design} \label{sec:Design}
This section will describe the design process for a prototype as a requirements-driven process. These requirements as given for the project are: {(a)} it will be launched from an approximately 3 inch tube (70-85mm), {(b)} it should fly ballistically to reach an altitude of 10m, {(c)} it should be able stabilize its flight after launch. In addition, {(d)} it should be a multirotor, and {(e)} it should be able to carry a payload of 200g.

From this set of requirements we can derive functional requirements that help the design process: the first requirement sets a form factor and, combined with requirement {(d)}, requires that the vehicle be able to deploy its arms that hold the motors. Requirement (a) also implies high vertical loads during launch, which will drive the structural design. Requirement {(e)} does not constrain the design space, as the vehicle is more volume limited than thrust limited. 
%
%

\begin{table}[htbp]
  \centering
  \caption{SQUID System Properties}
  
    \begin{tabular}{lr}
    \toprule
    Property & \multicolumn{1}{r}{Value} \\
    \midrule
    Mass  & 530 g \\
    Inertia about yaw axis, folded & $0.4\,10^{-3}$kg m$^2$ \\
    ~~~~"~~~~~"~~~~~~"~~~~"~~, unfolded & $2.3\,10^{-3}$kg m$^2$ \\
    Inertia about pitch axis, folded & $2.0\,10^{-3}$kg m$^2$ \\
    ~~~~"~~~~~~"~~~~~~"~~~~"~~, unfolded  & $1.6\,10^{-3}$kg m$^2$ \\
    Length & 270mm \\
    Folded Diameter & 83mm ($\approx$3in)  \\
    Maximum amperage & 38 A \\
    Thrust at hovering & 28\% \\
    Launch speed & 15m/s \\
    \bottomrule
    \end{tabular}%
  \label{tab:squidSummary}%
   \vspace{-0.3cm}
\end{table}%

This section will focus on the new challenges compared to a standard multirotor: first, the limited volume reduces the number of possible choices for most of the components. Second, the arms are not rigidly attached to the body. This will induce vibrations that affect the structure and control. Lastly, the strong vertical acceleration during launch imparts a large axial load on the multirotor. The main consequence of this high acceleration is the need to reinforce the structure, as well as ensure all components are properly secured and electrical connectors are tightly locked. Table~\ref{tab:squidSummary} provides a summary of the main design figures and Table~\ref{tab:SQUIDcomponents} contains a list of key SQUID components. \\

\begin{figure*}[htpb]
\centering
\tikzset{every picture/.style={line width=0.75pt}} 
\begin{tikzpicture}[x=0.75pt,y=0.75pt,yscale=-1,xscale=1]
\path (-10,300); 
\draw (244.75,145.44) node  {\includegraphics[width=217.13pt,height=217.41pt]{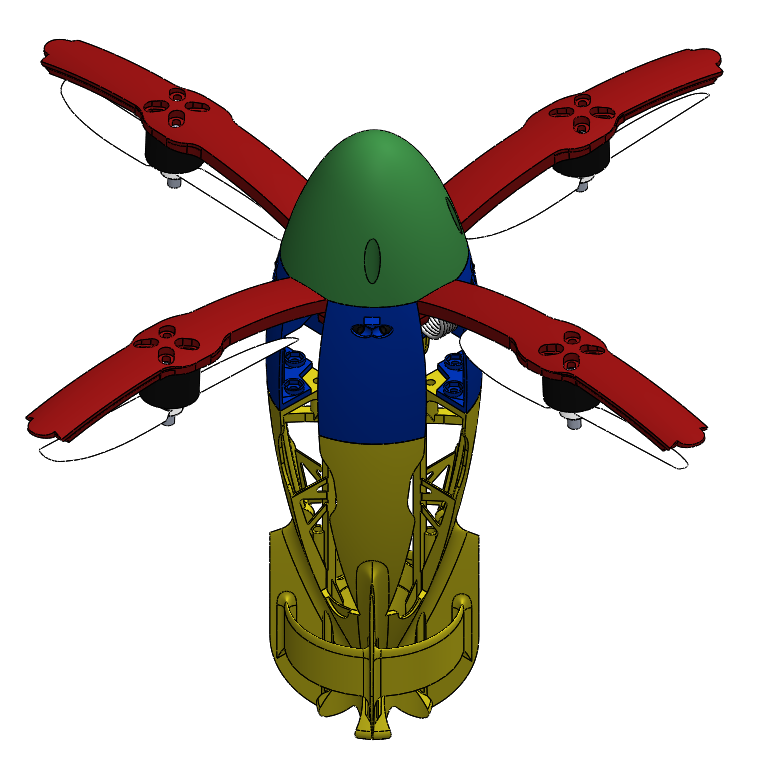}};
\draw (588.3,171.55) node  {\includegraphics[width=220.5pt,height=223.61pt]{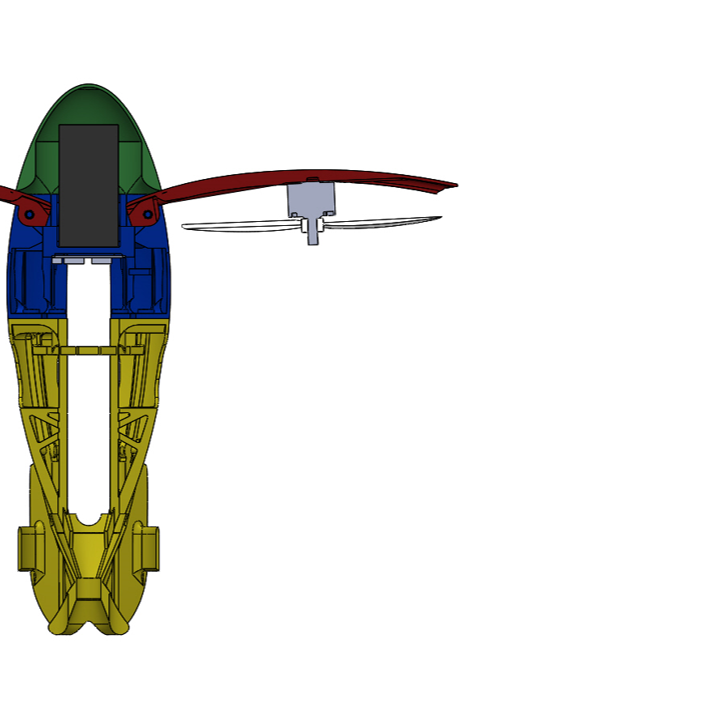}};
\draw    (180.24,264.11) -- (213.46,247.02) ;
\draw [shift={(215.24,246.11)}, rotate = 512.78] [fill={rgb, 255:red, 0; green, 0; blue, 0 }  ][line width=0.75]  [draw opacity=0] (8.93,-4.29) -- (0,0) -- (8.93,4.29) -- cycle    ;

\draw    (316.24,206.11) -- (281.73,175.43) ;
\draw [shift={(280.24,174.11)}, rotate = 401.63] [fill={rgb, 255:red, 0; green, 0; blue, 0 }  ][line width=0.75]  [draw opacity=0] (8.93,-4.29) -- (0,0) -- (8.93,4.29) -- cycle    ;

\draw    (235.24,28.11) -- (241.58,46.22) ;
\draw [shift={(242.24,48.11)}, rotate = 250.70999999999998] [fill={rgb, 255:red, 0; green, 0; blue, 0 }  ][line width=0.75]  [draw opacity=0] (8.93,-4.29) -- (0,0) -- (8.93,4.29) -- cycle    ;

\draw [dashed]   (487,91.33) -- (516.74,91.33) -- (516.74,126) -- (487,126) -- cycle ;
\draw    (516.74,91.33) -- (645,148.17) ;

\draw    (487,126) -- (535.3,243) ;

\draw (590.15,195.58) node  {\includegraphics[width=82.28pt,height=71.13pt]{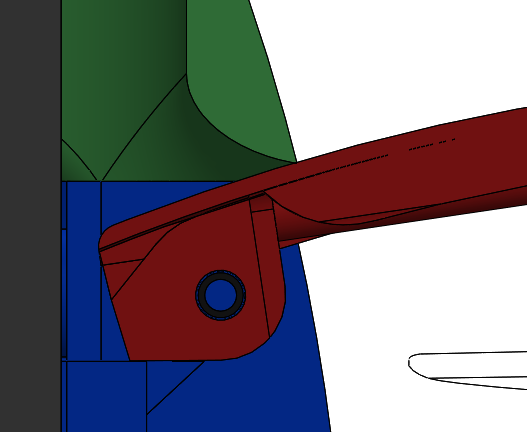}};
\draw (380.5,106.41) node  {\includegraphics[width=90pt]{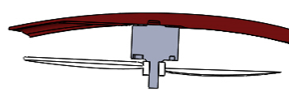}};
\draw  (535.3,148.17) -- (645,148.17) -- (645,243) -- (535.3,243) -- cycle ;
\draw    (445.17,73.33) -- (477.43,85.49) ;
\draw [shift={(479.3,86.2)}, rotate = 200.65] [fill={rgb, 255:red, 0; green, 0; blue, 0 }  ][line width=0.75]  [draw opacity=0] (8.93,-4.29) -- (0,0) -- (8.93,4.29) -- cycle    ;

\draw    (440.17,162.33) -- (481.55,139.18) ;
\draw [shift={(483.3,138.2)}, rotate = 510.77] [fill={rgb, 255:red, 0; green, 0; blue, 0 }  ][line width=0.75]  [draw opacity=0] (8.93,-4.29) -- (0,0) -- (8.93,4.29) -- cycle    ;

\draw (65.12,158.29) node  {\includegraphics[width=73.32pt,height=184.93pt]{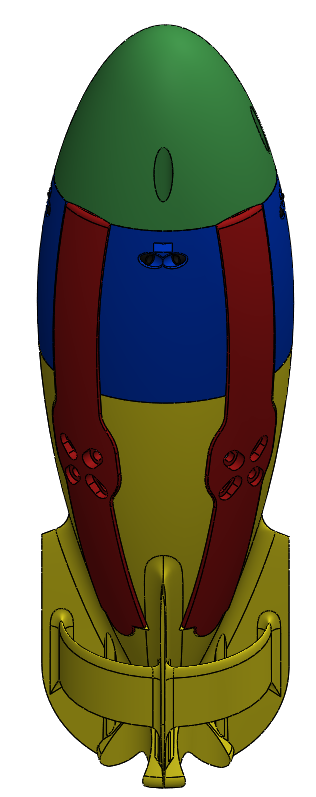}};

\draw (145,264) node  [align=left,scale=0.9] {fin ring};


\draw (350,222) node  [align=left,scale=0.9] {aerodynamic\\shell};
\draw (223,20) node  [align=left,scale=0.9] {nose};
\draw (420,66) node  [align=left,scale=0.9] {battery};
\draw (405,167) node  [align=left,scale=0.9] {electronics};
\draw (588,255) node  [align=left,scale=0.9] {hinge zoom};
\end{tikzpicture}
\caption{SQUID CAD model. From left to right: ballistic configuration view, multirotor configuration view and section view with a hinge closer look.}
\label{fig:cad}
\end{figure*}
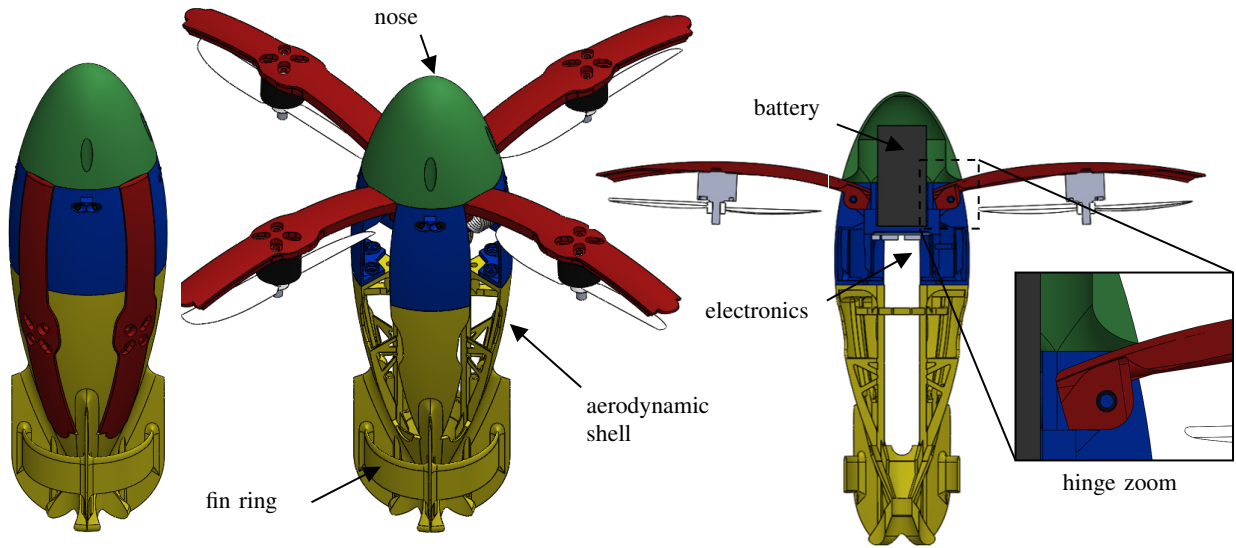

\noindent \textbf{Vehicle Sizing and Aerodynamic Design:} Due to the launcher diameter constraint, we design the outer shell in a compromise of internal volume, air drag, and stability (see Figure \ref{fig:cad} for the selected shape). No detailed numerical simulations were performed, but we followed the insights from classic projectile design~\cite{hoerner1958fluid,jorgensen1973prediction} with aerodynamic forces and moments estimated as:

\begin{figure}[thpb]
	\centering
	\tikzset{every picture/.style={line width=0.75pt}} 
	
	\begin{tikzpicture}[x=1pt,y=1pt,yscale=1,xscale=1]

	\draw (60,0) node  {\includegraphics[width=120pt]{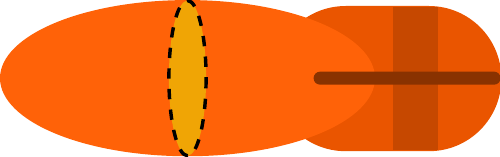}};

	\draw (-25,-20) node  [align=center] {$V$};
	\draw [->, ultra thick] (-60,-20) -- (0,-2) ;
	
	\draw (-30,10) node  [align=center] {$v_a$};
	\draw [->,dashed] (-62,2) -- (0,2) ;
	
	\draw (-70,-10) node  [align=center] {$v_n$};
	\draw [->,dashed] (-62,-20) -- (-62,0) ;
	
	\draw (75,35) node  [align=center] {$d, A_\text{front}$};
	\draw (45,0) -- (75,30);
	\draw [fill] (45,0) circle [radius=2];

	\draw (30,35) node  [align=center] {$\volume, A_\text{side}$};
	\draw (25,0) -- (30,30);
	\draw [fill] (25,0) circle [radius=2];	

	\draw (120,35) node  [align=center] {$A_\text{fin}$};
	\draw (100,0) -- (120,30);
	\draw [fill] (100,0) circle [radius=2];

	\draw (5,30) node  [align=center] {$x$};
	\draw [->] (0,20) -- (0,40);
	\draw (-10,15) node  [align=center] {$z$};
	\draw [->] (0,20) -- (-20,20);
	\draw [fill] (0,20) circle [radius=1];	
	
	\draw (60,-27) node [align=center] {$L$};
	\draw (0,-17) -- (0,-22) -- (120,-22) -- (120,-17);

	\end{tikzpicture}
	\caption{Aerodynamic Nomenclature}
	\label{fig:forces}
\end{figure}
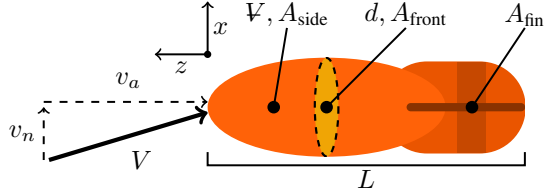

\begin{align}
M_\text{munk} &= \rho v_a v_n \volume (1-d/L) \\
F_{\text{base},n} &= \rho v_a v_n A_\text{front} C_{d,\text{front}} \\
F_{\text{lift},n} &= \frac{1}{2} \rho v_a v_n A_\text{fin} C_{l\alpha,\text{fin}}\\
F_{\text{side},n} &= \frac{1}{2} \rho v_n v_n A_\text{side}  C_{d,\text{side}}
\end{align}

Where $F_{\text{base},n}$, $F_{\text{lift},n}$, and $F_{\text{side},n}$ are the components of the base drag, fin lift, and side drag taken normal to the primary axis of the body, and $M_\text{munk}$ is the Munk moment. Symbols $\rho$, $v_a$, $v_n$, $L$, $d$, and $\volume$ are the air density, axial and normal velocities, length, diameter, and volume respectively. Equations 1-4 are applicable for the designed SQUID model (a mildly streamlined body operating beyond turbulent transition) \cite{hoerner1958fluid}, but are not expected to apply to substantially smaller, slower, or smoother aircraft that may be more Reynolds-sensitive. The aerodynamic center, which should be placed after the center of mass for passive stability, is given by:

\begin{align}
z_\text{AC} = \frac{-M_\text{munk}+F_{\text{base},n}L+F_{\text{fin},n}L+F_{\text{side},n}L/2}{F_{\text{base},n}+F_{\text{lift},n}+F_{\text{side},n}}
\end{align}

The Munk moment is unstable and grows with the object's volume, while both the drag and fin lift are generally stabilizing. Accordingly, both standard fins and a ring-fin are required to lower the aerodynamic center (and increase fin structural integrity) to compensate for the low-drag high-volume design. The estimated aerodynamic center location of the final design resides at roughly 65\% of the folded SQUID length, leading to stable damped pitch oscillations of 0.6s period and stability margin of 5cm. 

The arm deployment has three effects related to aerodynamic stability: it moves the center of mass 3cm towards the nose (increasing stability), it increases both the axial and normal drag (increasing damping but also shifts the aerodynamic center 3cm towards the nose due to the arm location), and it increases the yaw inertia by a factor of 5 (decreases yaw rate due to conservation of angular momentum). The net effect maintains stability during the transition to flight geometry. Deliberate spin-stabilization during launch was rejected for ease of piloting and to simplify the transition dynamics between launch and flight. The design was experimentally validated as is shown in Section~\ref{sec:Testing}.  \\ 

\noindent \textbf{Propeller and motor selection:} the next step is to select the electrical components. The propeller size can be derived for ideal disc loading at hover \cite{leishman2006principles}:

\begin{align}
\frac{mg}{4\pi r^2_\text{prop}} = \frac{1}{2} \rho v^2_\text{tip} (\sigma_\text{prop} C_{d0, \text{prop}}/k_\text{prop})^{2/3}
\end{align}

Where $\sigma_\text{prop}\approx 0.1$, $C_{d0, \text{prop}}\approx 0.02$, $k_\text{prop}\approx 1.25$ are rough estimates of the propeller solidity, nominal drag coefficient, and induced loss factors. Assuming a tip speed of $v_\text{tip} = 100$ m/s at hover (Mach 0.3), the ideal propeller size for hover with payload is around 6 or 7 inches. However, given the strong volume constraints for a passively stable aeroshell that folds within the launch tube, we can only choose the biggest propeller accommodated in the full system design, in this case 5 inches in diameter. This still gives us a large margin of excess thrust for operations using racing motors designed for smaller propellers. Knowing the propeller size, we select the motor Air40 from TMotors as it can drive this propeller and it has a good compromise of responsiveness and efficiency. Note that, despite the fact that flight time is not a requirement for this vehicle and therefore the design is not optimized for it, the battery was selected as the biggest battery that can be accommodated in the given space, in this case a Tattu 850mAh. \\ 

\noindent \textbf{Component Placement:} The heaviest component, the battery, is placed as close to the nose as possible to increase the center of mass vertical location. This will increase aerodynamic stability during the ballistic launch \cite{hoerner1958fluid}. The rest of the electronic components are placed directly below the battery: autopilot, BEC and radio receiver. In addition, the ESC are placed on each arm to avoid the limited space on the core and the radio antennas are extended to the bottom core piece for improved radio signals. Similarly, the GPS module is situated on top of the battery for better coverage. \\ 

\noindent \textbf{Structure Design:} The main structural load for SQUID is due to the vertical acceleration from launch. From early experiments, we measured a vertical acceleration of 50G's (490 m/s\textsuperscript{2}) to meet the height requirement with a sub-meter acceleration distance. This acceleration will appear as a volumetric force to all components. In particular, we designed the main structure to connect the inertial load from the battery, situated at the top and the heaviest component, to the launcher at the bottom. The 3D printed parts were printed using high impact resistance materials, using the Markforge printer with Onyx and carbon fiber. Another important load is due to arm unfolding. Limited space prevents us from adding additional material to make the arms more rigid, and the curved surface limits the use of traditional CNC methods. Another benefit of 3D printed carbon fiber is the added rigidity, which is needed in our design in order to provide a tight fit when the arms are folded. \\

\noindent \textbf{Hinge Design:} The hinges allow the arms to rotate freely after release and limit their movement so that the propellers are horizontal during normal flight. The unfolding limit is set by a mechanical stop. The hinges each hold a torsion spring that push the arms to open after their release. During normal flight, the springs are strong enough to maintain open the arms and provide resistance against vertical disturbances. An overly stiff spring creates large shock loads during arm unfolding. During launch, the arms fold to slightly beyond 90$^\circ$ from their open posture so that the propellers are tilted inside the body to allow more space at the top for the electronics. \\

\begin{figure}[thpb]
	\centering
\tikzset{every picture/.style={line width=0.75pt}} 

\begin{tikzpicture}[x=0.75pt,y=0.75pt,yscale=-0.7,xscale=0.7]

\draw (127.57,94.7) node  {\includegraphics[width=135pt,height=86pt]{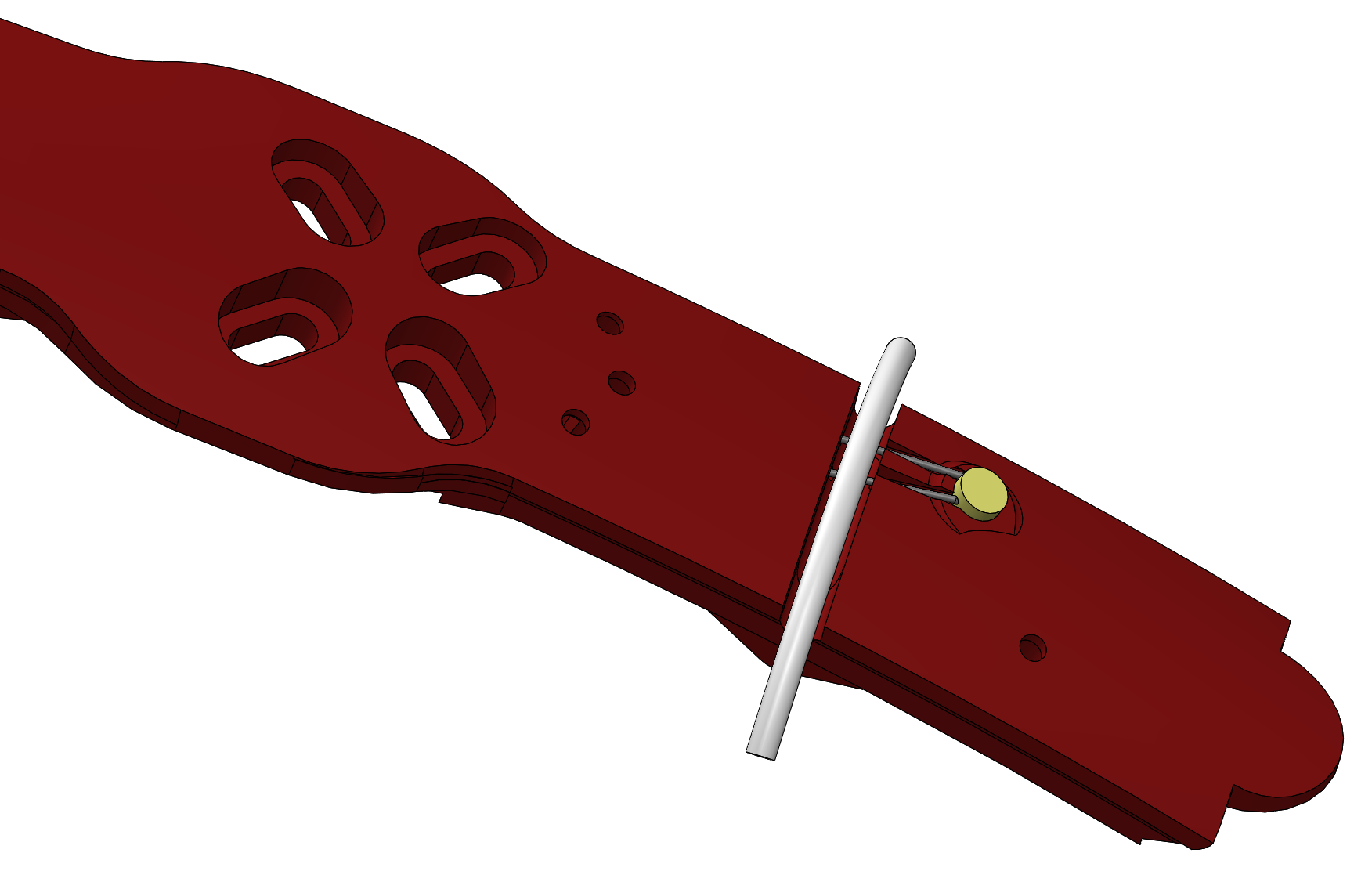}};
\draw    (209,43) -- (177,89.79) ;
\draw [shift={(175,91.43)}, rotate = 304.96] [fill={rgb, 255:red, 0; green, 0; blue, 0 }  ][line width=0.75]  [draw opacity=0] (8.93,-4.29) -- (0,0) -- (8.93,4.29) -- cycle    ;

\draw    (85,143) -- (133.14,145.5) ;
\draw [shift={(140.14,145.43)}, rotate = 537.95] [fill={rgb, 255:red, 0; green, 0; blue, 0 }  ][line width=0.75]  [draw opacity=0] (8.93,-4.29) -- (0,0) -- (8.93,4.29) -- cycle    ;

\draw (200,35) node  [align=left] {Nichrome wire};
\draw (10,141) node  [align=left] {Monofilament line};
\end{tikzpicture}
\caption{Release Mechanism Detail.}
	\label{fig:release}
\end{figure}
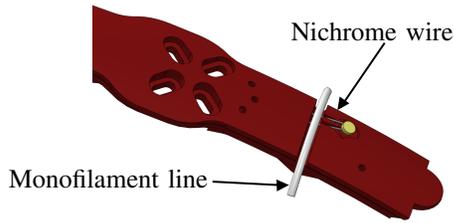

\noindent \textbf{Release Mechanism:} While several potential release mechanisms were considered, including designs employing electromagnets and servo motors, we selected a nichrome burn-wire trigger due to its reliability, efficient use of space, low susceptibility to G-forces, and low mass. Current passing through the nichrome wire causes it to heat up and cut through a restraining loop of nylon monofilament line. This technique has been previously used on CubeSats, proving effective in both Earth atmosphere and vacuum~\cite{bib:nichrome}. The greatest downside of a nichrome release mechanism is the inconvenience of manually replacing the monofilament line after every launch, so the mechanism was designed for ease of access. A shallow groove runs around the circumference of the SQUID in its ballistic configuration to hold a loop of monofilament line in place. The tension in the arms causes them to push outwards against the line, but the chosen line is strong enough to withstand both the spring and launch forces without snapping. Mounted on one of the arms is a length of nichrome wire, held under tension by screw terminals that have been heat-set into the arm. The nichrome wire presses against the line, so that when heated it severs the line and releases the spring-loaded arms. \\ 

\begin{table}[htbp]
\vspace{-0.1cm}
  \centering
  \caption{Key SQUID components}
    \begin{tabular}{lrrr}
    \toprule
    Component & \multicolumn{1}{l}{Name} & \multicolumn{1}{l}{Weight (g)} & \multicolumn{1}{l}{Quantity} \\
    \midrule
    Autopilot & \multicolumn{1}{l}{Pixracer running PX4} & 14  & 1 \\
    Motor & \multicolumn{1}{l}{T-Motors Air40} & 24  & 4 \\
    ESC   & \multicolumn{1}{l}{T-Motors F30A} & 7     & 4 \\
    Propeller & \multicolumn{1}{l}{DAL 5050} & 4     & 4 \\
    Receiver & \multicolumn{1}{l}{FrSky R-RXR} & 1.2   & 1 \\
    Battery & \multicolumn{1}{l}{Tattu 850mAh} & 104   & 1 \\
    Power board & \multicolumn{1}{l}{ACSP7} & 15    & 1 \\
    Frame & \multicolumn{1}{l}{Custom}      & 181   & 1 \\
    Arms &  \multicolumn{1}{l}{Custom}     & 16 & 4 \\
    \bottomrule
    \end{tabular}%
  \label{tab:SQUIDcomponents}%
\end{table}%

\section{Operations} \label{sec:Operations}
\begin{figure}[hptb]
\vspace{-0.1cm}
\tikzset{every picture/.style={line width=0.75pt}} 

\begin{tikzpicture}[x=0.75pt,y=0.75pt,yscale=-1,xscale=1]

\draw (68.91,242.7) node [rotate=-20] {\includegraphics[width=71.15pt,height=41.26pt]{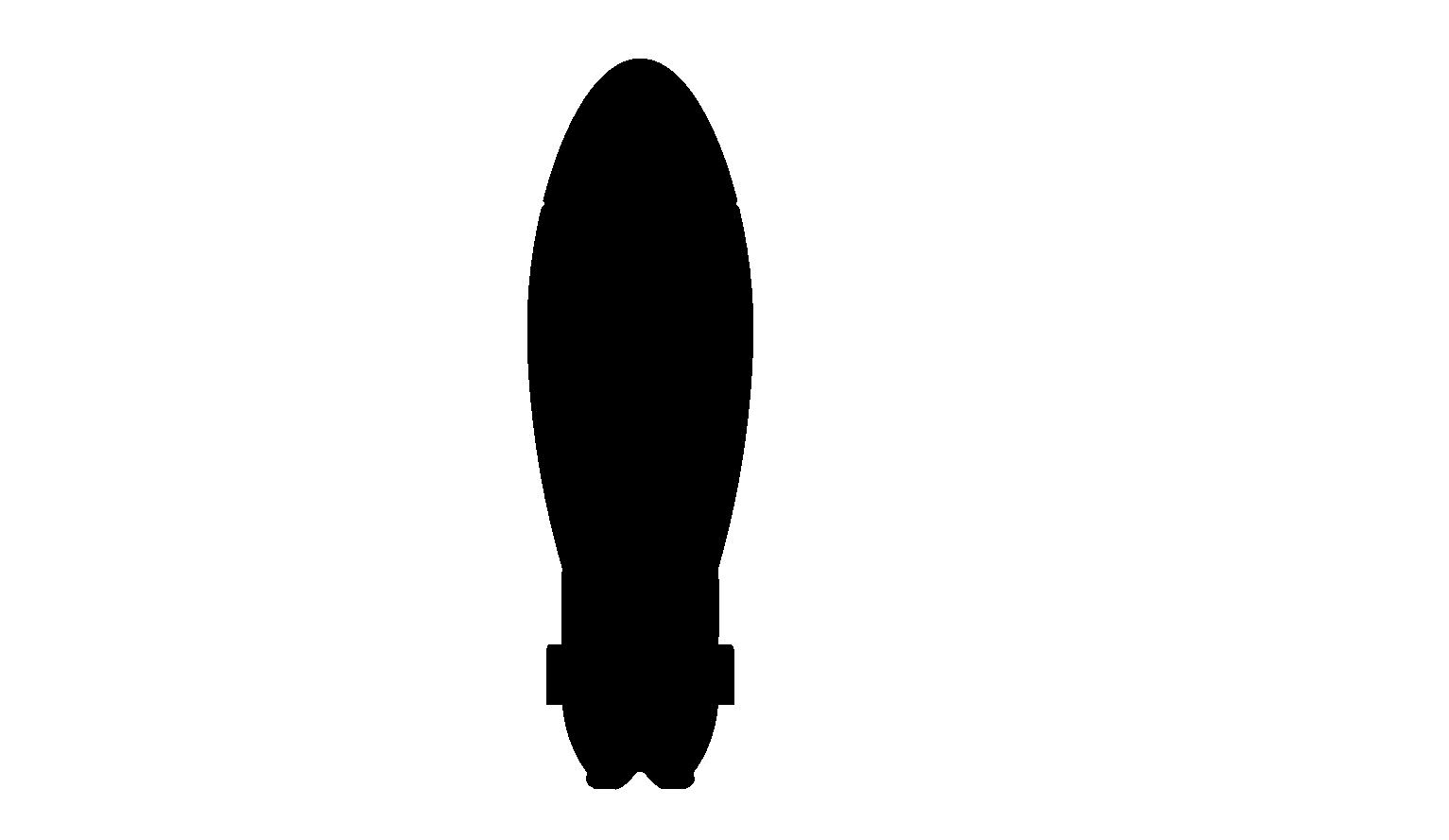}};
\draw (98.32,175.47) node [rotate=-20] {\includegraphics[width=73.43pt,height=42.58pt]{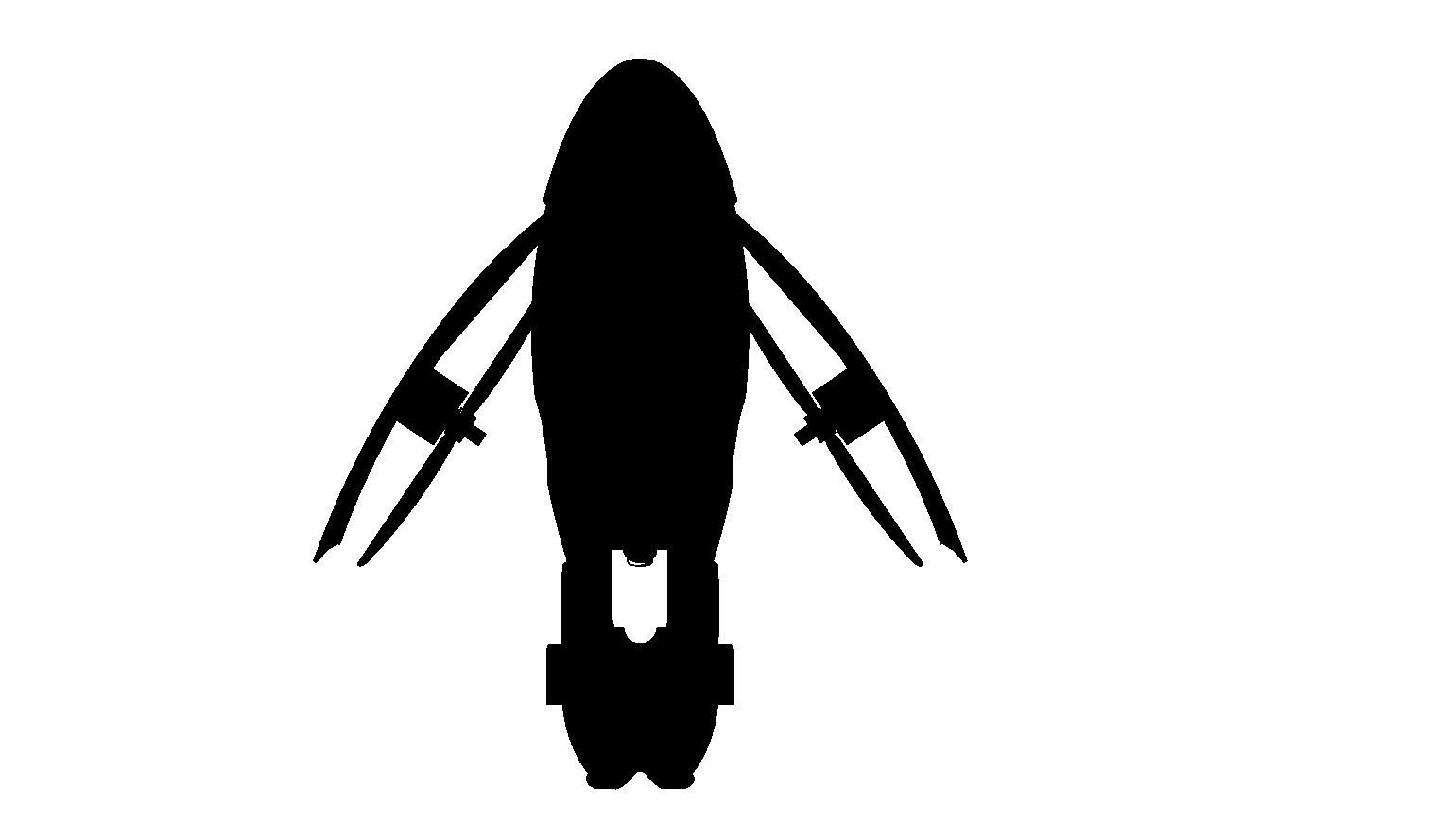}};
\draw (135.41,92.48) node [rotate=-10] {\includegraphics[width=73.43pt,height=42.58pt]{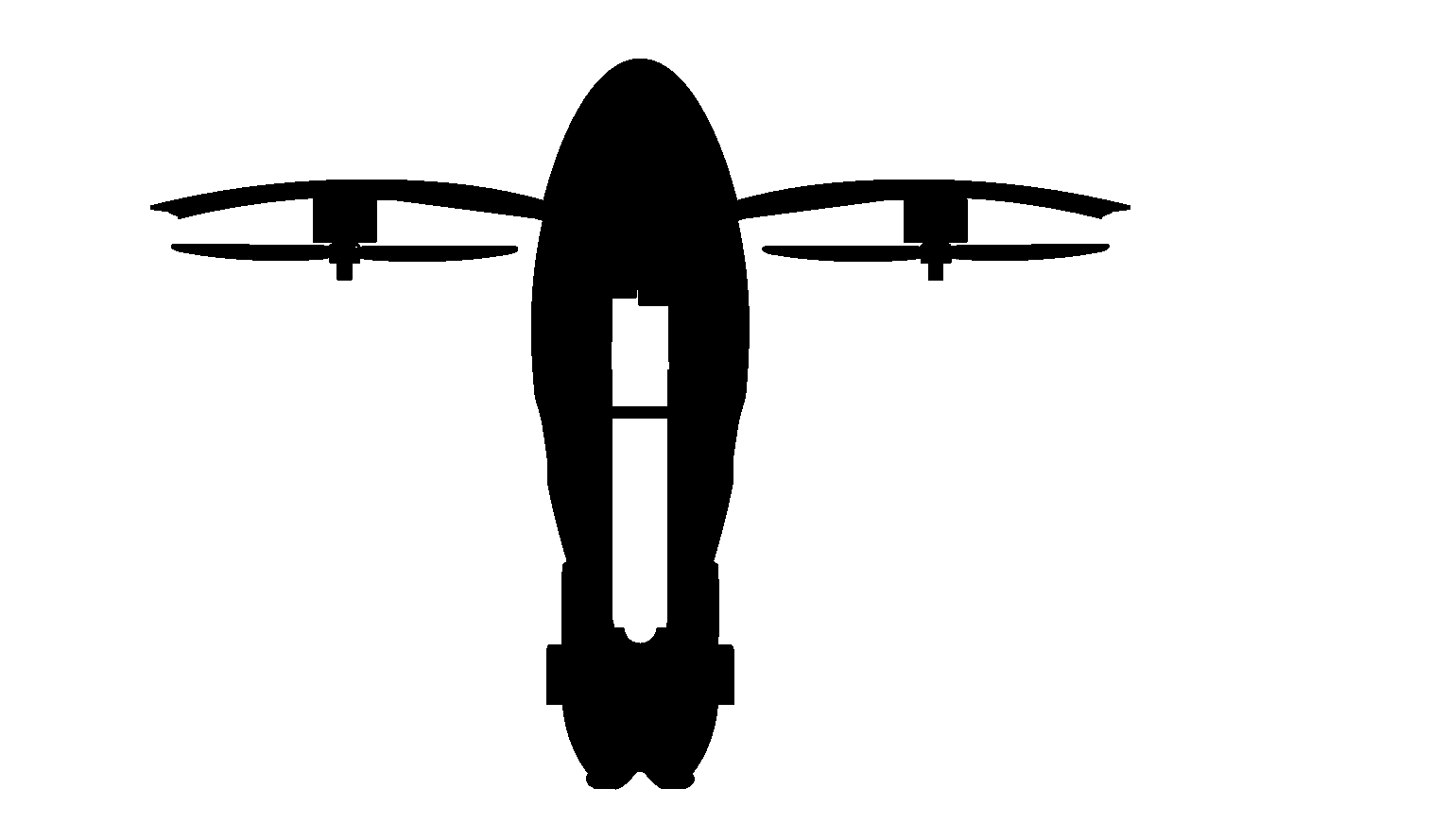}};
\draw (229.08,35.48) node  {\includegraphics[width=73.43pt,height=42.58pt]{front_view_open_black.PNG}};
\draw  [fill={rgb, 255:red, 0; green, 0; blue, 0 }  ,fill opacity=1 ] (37.87,278.83) -- (54.39,284.84) -- (44.2,312) -- (25.87,312) -- cycle ;
\draw (180,180) node [scale=0.9] [align=left] {(4) Arms deployment};
\draw (155,240) node [scale=0.9] [align=left] {(3) Unpowered flight};
\draw (190.96,120) node [scale=0.9] [align=left] {(5) Stabilization};
\draw (160,300) node [scale=0.9] [align=left] {(1-2) Resting and launch inside the barrel};
\draw (240,80) node [scale=0.9] [align=left] {(6) Standard multirotor \\ \ \ \ controlled flight};
\end{tikzpicture}
	\caption{SQUID deployment sequence}
	\label{fig:deploymentSequence}
\end{figure}

The operation of SQUID is composed of six different phases from loading to controlled flight. See Figure~\ref{fig:deploymentSequence} for an illustrative diagram.

\begin{figure}[thpb]
  \centering
  \includegraphics[width=0.4\linewidth]{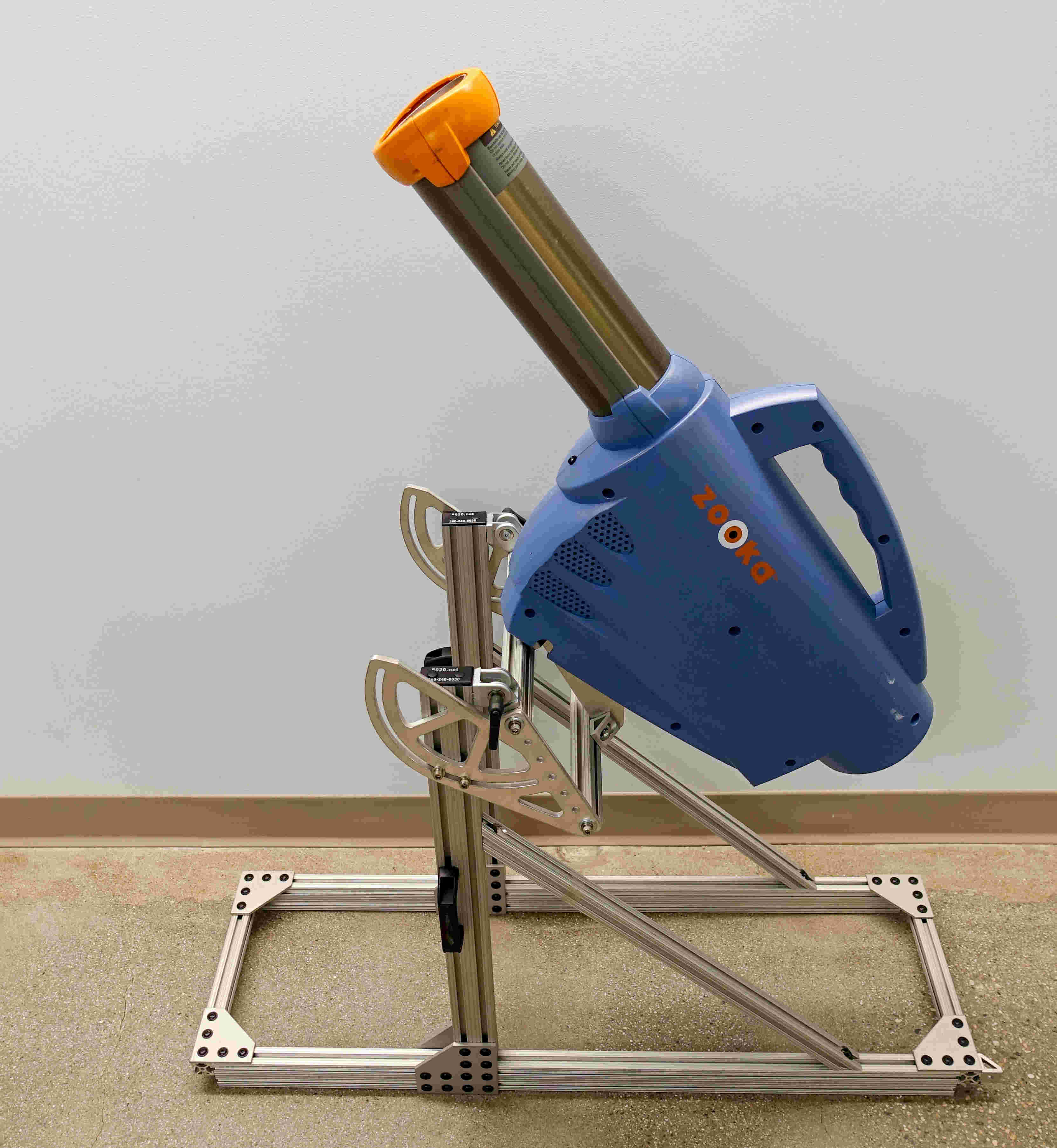}
  \caption{The pneumatic baseball pitching machine used to launch the SQUID prototype.}
  \label{fig:launcher}
\end{figure}

\begin{enumerate}
\item \textbf{Resting inside the launching device:} The vehicle is static and ready to be launched. Before this phase, the vehicle has been turned on and armed. In order to keep compatibility with the rest of the PX4 stack, the vehicle is set to \textit{'kill'} mode in order to neglect all input commands.
\item \textbf{Acceleration inside the barrel:} After launch is triggered, the compressed air accelerates the vehicle through a 76cm barrel with high \textit{g} forces. This acceleration can be used by the autopilot to detect the launch. Figure~\ref{fig:acc} shows a typical acceleration profile throughout operation. The first acceleration spike corresponds to the launch acceleration. We use a pneumatic ZS740 baseball pitching machine from Zooka (see Figure~\ref{fig:launcher}), which can realize 15m/s (35mph) muzzle velocity for the described SQUID prototype.
\item \textbf{Unpowered flight:} After launch, SQUID travels at high speeds and follows a parabolic trajectory. In the case of a moving vehicle launch, SQUID's relative velocity is the composition of the launch speed and the moving vehicle speed.
\item \textbf{Arms deployment:} The folded arms are initially retained by the monofilament line. They open when a relay actuates the nichrome burn wire. Without the monofilament line, the torsion spring deploys the arms. While arm deployment angle is not controlled, the arms fully deploy in 70ms, but they recoil by up to 30$^\circ$ before the motion is damped.
\item \textbf{Stabilization:} The pilot sends the command to \textit{'unkill'} the drone and it automatically orients itself to the hovering attitude. For convenience, in the current SQUID prototype, the pilot must compensate for altitude and lateral motion, but the vehicle includes a GPS for waypoint navigation. Future versions are being designed to provide autonomous stabilization using vision-based methods, which requires a larger volume to house a computer vision camera, 1D lidar and a bigger on-board computer. In \cite{faessler2015automatic}, the authors implements an algorithm to recover midair using on-board sensors. SQUID requires a similar approach but the speeds are much greater.
\item \textbf{Standard Multirotor Controlled Flight:} After SQUID stabilizes, it operates as a normal multirotor. The current design was not optimized for long battery life, but future prototypes might be able to carry different batteries depending on the mission length. While SQUID does not have dedicated landing legs, it can safely land if the bottom touches the ground first at a low speed. It naturally falls to one side without damaging any component. Another landing method is to grab the bottom part of SQUID, as shown in the accompanying video.
\end{enumerate}

\section{Field Testing} \label{sec:Testing}
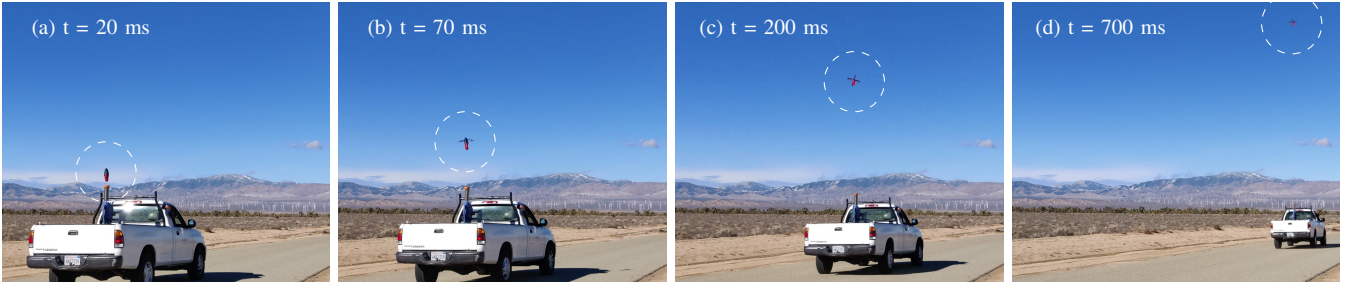
\begin{figure*}[htpb]
  \centering
\begin{tikzpicture}
    \draw (0, 0) node[inner sep=0] {\includegraphics[width=0.245\linewidth, trim={850 0 1450 850},clip]{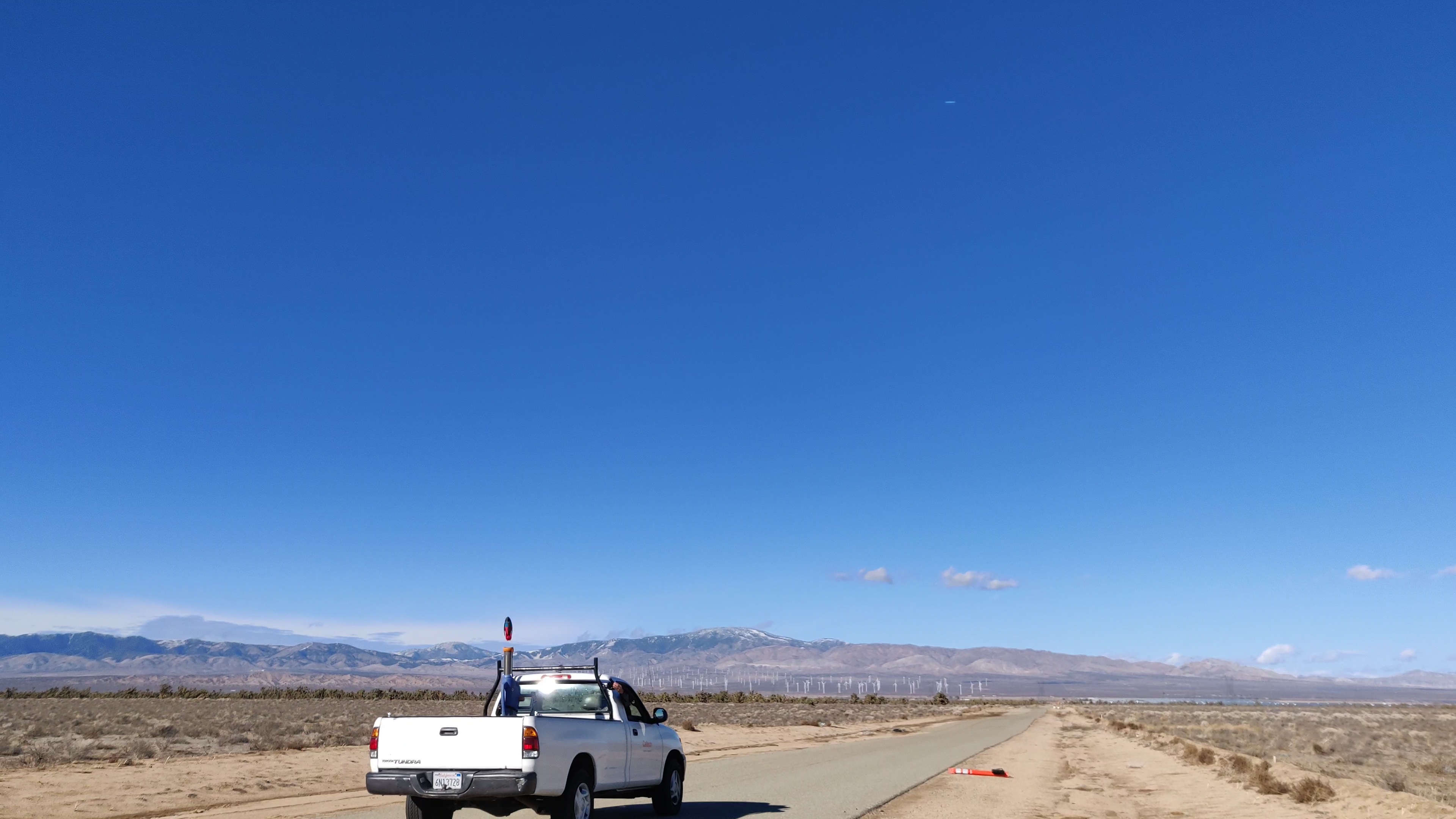}};
    \draw  (-1, 1.5) node [text=white,scale=0.9] {\small (a) t = 20 ms};
    \draw[white,dashed] (-0.8,-0.4) circle (0.4cm);
\end{tikzpicture}
\begin{tikzpicture}
    \draw (0, 0) node[inner sep=0] { \includegraphics[width=0.245\linewidth, trim={850 0 1450 850},clip]{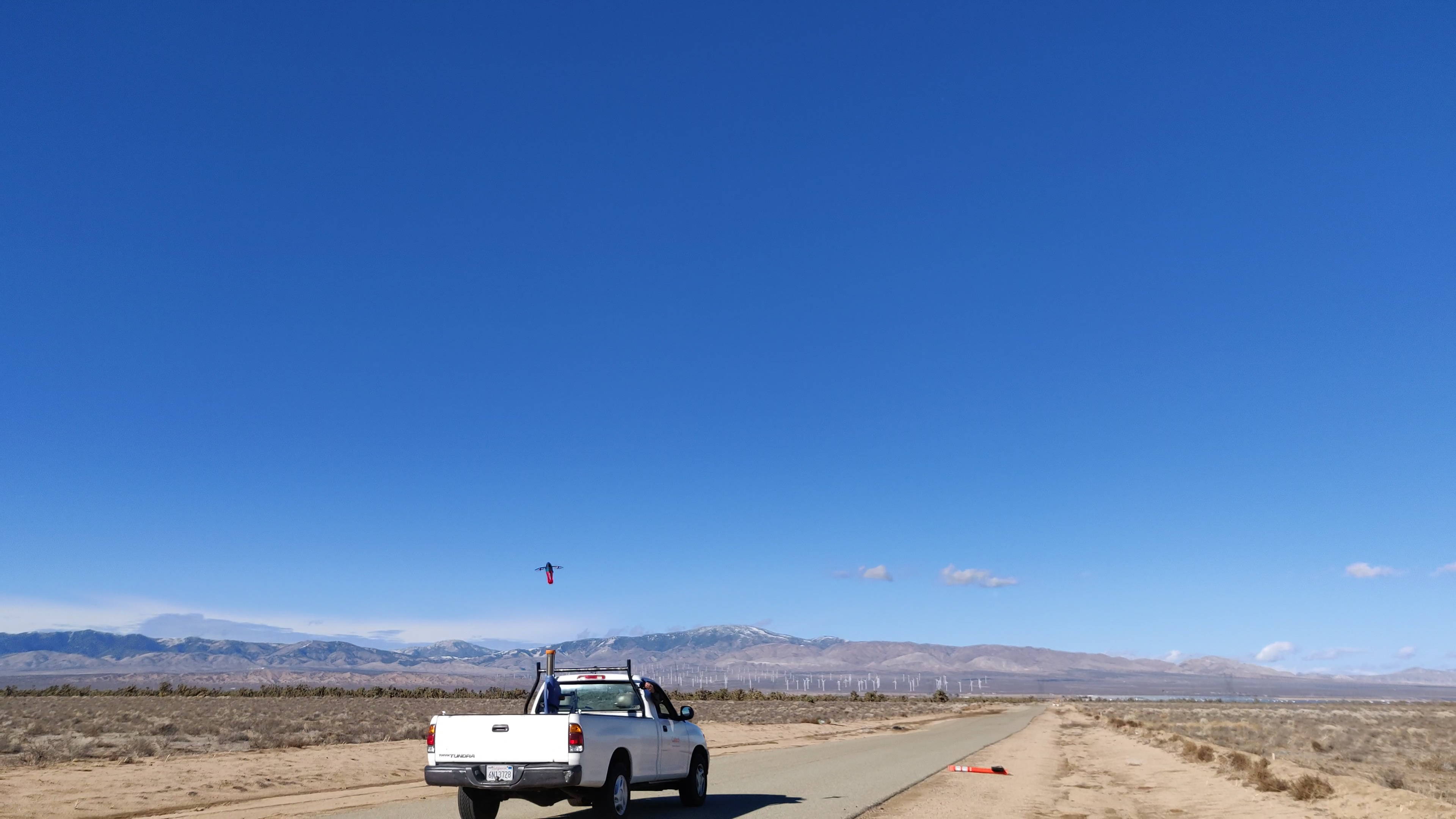}};
    \draw  (-1, 1.5) node[text=white,scale=0.9] {\small (b) t = 70 ms};
    \draw[white,dashed] (-0.5,0) circle (0.4cm);
\end{tikzpicture}
\begin{tikzpicture}
    \draw (0, 0) node[inner sep=0] {\includegraphics[width=0.245\linewidth, trim={850 0 1450 850},clip]{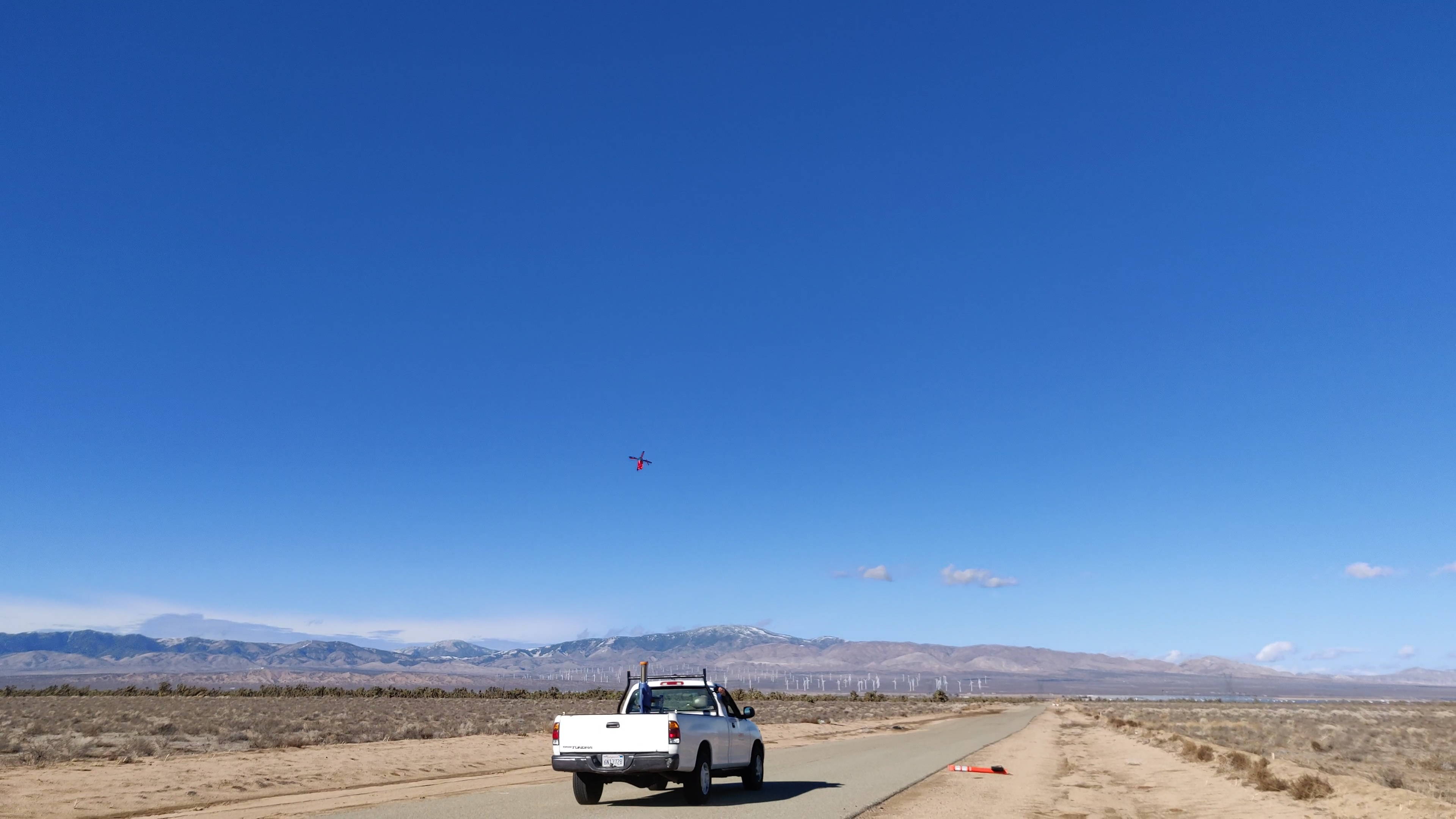}};
    \draw  (-1, 1.5) node[text=white,scale=0.9] {\small (c) t = 200 ms};
    \draw[white,dashed] (0.2,0.8) circle (0.4cm);
\end{tikzpicture}
\begin{tikzpicture}
    \draw (0, 0) node[inner sep=0] {\includegraphics[width=0.245\linewidth, trim={850 0 1450 850},clip]{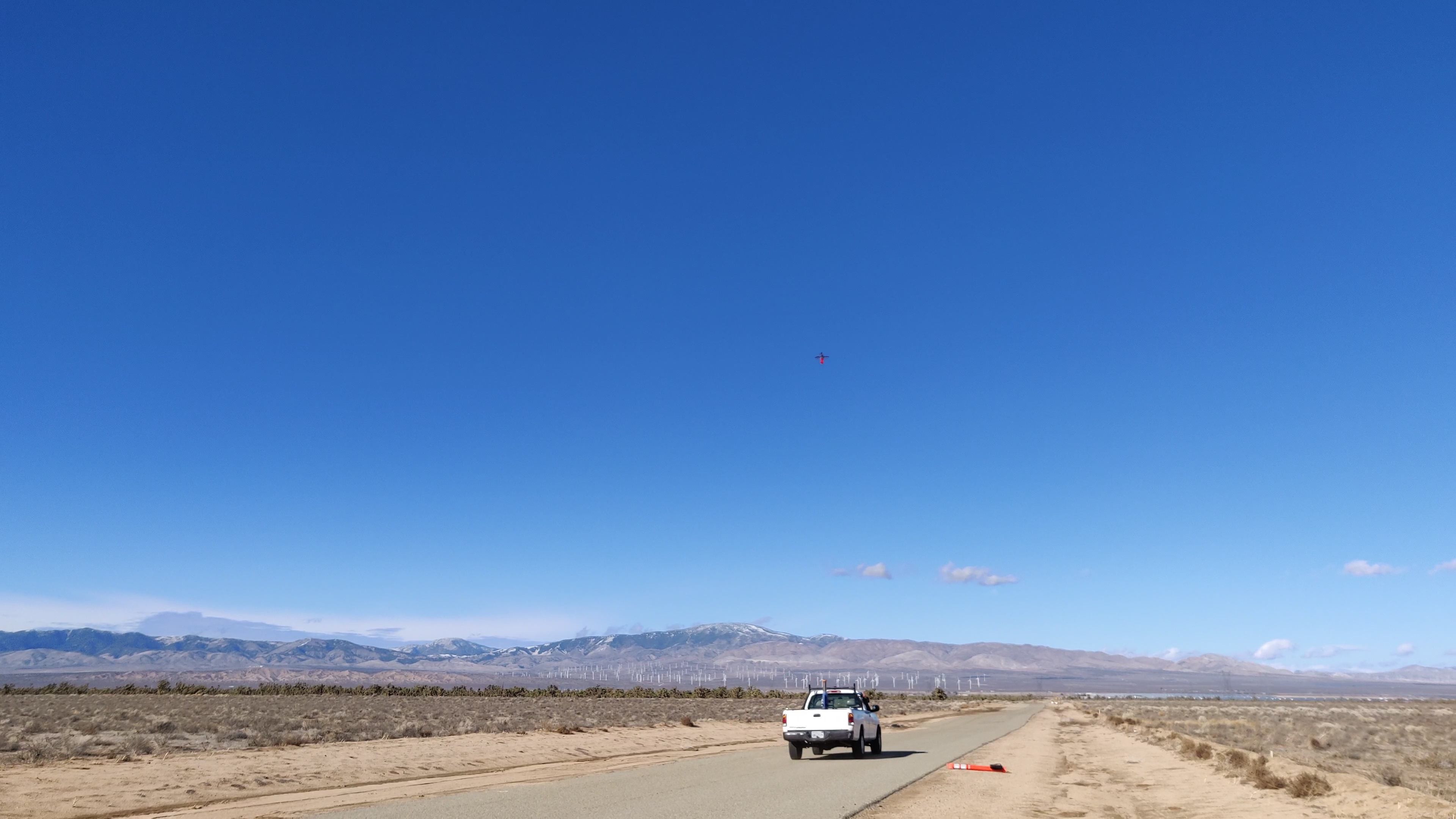}};
    \draw  (-1, 1.5) node[text=white,scale=0.9] {\small (d) t = 700 ms};
    \draw[white,dashed] (1.54,1.57) circle (0.4cm);
\end{tikzpicture}
  \caption{Launching from a moving vehicle snapshots with the time from launch on the upper left. From left to right: (a) 20ms after deployment the arms are still closed. It is moving straight up in the cannon direction. (b) The arms have been deployed around 70ms after launch. The vehicle is still moving up. (c) The body is passively aerodynamically stable so it predictably orients itself against its relative velocity. By 200ms it is oriented upwind. (d) In this snapshot the vehicle is already stable and hovering.}
  \label{fig:collage_deployment}
\end{figure*}

\begin{figure}[thpb]
  \centering
  \includegraphics[width=1.0\linewidth]{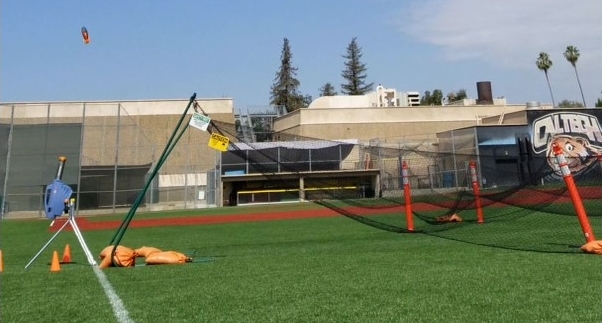} 
  \caption{The field testing setup on the Caltech sports field with a net to protect the SQUID prototype from crashes}
  \label{fig:field}
\end{figure}

\begin{figure}[thpb]
  \centering
  \includegraphics[width=0.90\linewidth]{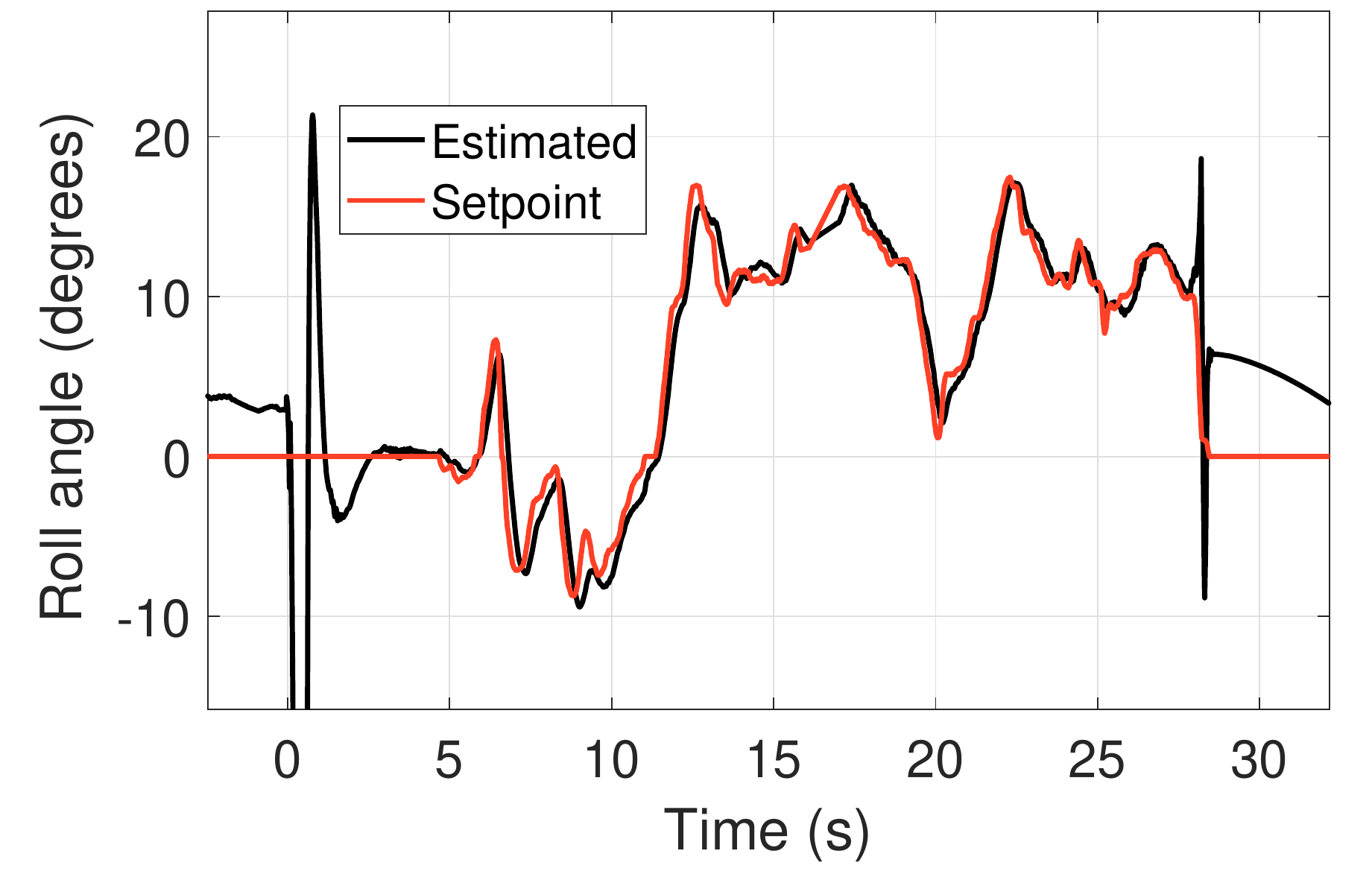}
  \caption{Roll angle profile during the moving vehicle test. The vehicle takes around one second to stabilize to its roll command.}
  \label{fig:roll}
\end{figure}

\begin{figure}[thpb]
  \centering
  \includegraphics[width=0.9\linewidth]{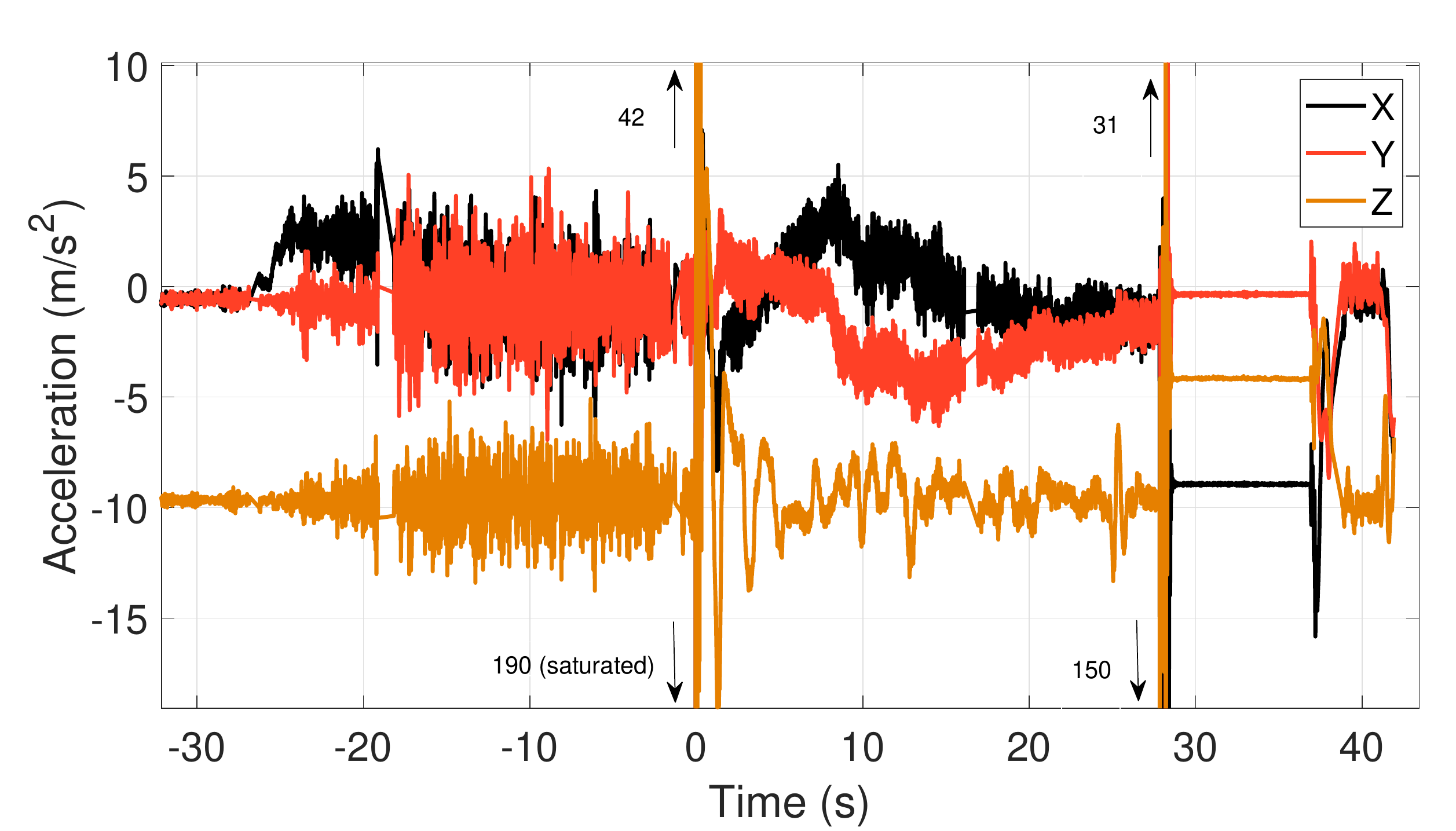}
  \caption{Acceleration profile during the moving vehicle test, where the \textit{x} axis is pointing forward, the \textit{z} axis is pointing up, and time starts when SQUID is launched. At -25s before launch, the vehicle accelerates to 80km/h (50mph) which can be seen as a constant acceleration on the $x$ axes. After that, the acceleration is very noisy due to the bumpy road. The launch is indicated by a large acceleration spike in the $z$ axis. There is another spike 29s later when the vehicle lands and tilts sideways onto its arms. During the flight, the $z$ acceleration is close to negative one-\textit{g} (9.8 m/s\textsuperscript{2}) indicating level flight, and the $x$ and $y$ acceleration commanded by the pilot compensate for the initial 50mph vehicle speed.}
  \vspace{-0.6cm}
  \label{fig:acc}
\end{figure}

We designed a set of tests to verify SQUID's capabilities. Here we will discuss the three main stepping stones during development. See the accompanying video for more details about field testing. 
\begin{enumerate}
\item \textbf{Aerodynamic test:} We used a mass model in order to evaluate aerodynamic effects in the vehicle prior to integrating electrical components, slowly increasing the fin size within volume constraints until enough stability margin was achieved for the test conditions. The selected shape includes a ring-fin for added stability and structural integrity. 
\item \textbf{Delayed deployment test:} This test demonstrates deployment from a static launcher, see Figure~\ref{fig:field} for a picture during midair flight. It contains all the phases described in Section~\ref{sec:Operations}. 
\item \textbf{Moving vehicle test:} On this test we launched SQUID from a car moving at 22m/s (50mph), see Figure~\ref{fig:car_video_frames}~and~\ref{fig:collage_deployment} for keyframes from the video and Figures~\ref{fig:acc}~and~\ref{fig:roll} for key data during flight. It demonstrates that SQUID can be deployed at high speeds without problems. 
\end{enumerate}

\section{Scaling Arguments} \label{sec:Scaling}
When designing a ballistic launch for a different-sized SQUID (larger tube diameter, etc.), the following non-dimensionalized argument can be used to predict the aerodynamic performance. This analysis broadens the scope of our field testing conclusions, which can then be applied to other aircraft given the appropriate scaling.

The launch trajectory of the multirotor must be a function of an input variable set; namely the launch velocity ($U$), vehicle velocity ($U_\text{vehicle}$), air properties (density and viscosity $\rho$ and $\mu$), gravity ($g$), time ($t$), and the geometry of the aircraft (mass $m$, diameter $d$, length $L$, inertia $I$). Given that these input variables can be expressed using three independent physical units (mass, time, and length), we can describe the same equations using three fewer non-dimensional variables than input variables. The following non-dimensional variables accordingly span the input space:
 \begin{gather}
 \tilde{t} = \frac{tU}{L},  \,\,\,\,  \textit{Fr} = \frac{U}{\sqrt{g L}}, \,\,\,\, \textit{Re} = \frac{\rho U L}{\mu}, \\
 \tilde{U}_\text{vehicle} = \frac{U_\text{vehicle}}{U}, \,\,\,\, \tilde{m} = \frac{m}{\rho L^3}, \,\,\,\, \tilde{d} = \frac{d}{L}, \,\,\,\, \tilde{I} = \frac{I}{\rho L^5} 
 \end{gather}
 
 Where $\textit{Fr}$ is the Froude number and $\textit{Re}$ is the Reynolds number. Further nondimensional groups can represent the fin area ratio $A_\text{fin}/L^2$ etc. and other geometry details, but are generally held consistent for exact scale models. Reynolds number $\textit{Re}$ effects are expected to be minimal and can be neglected for models scaled by a single order-of-magnitude, as drag coefficients are only weakly dependent on $\textit{Re}$ given the fully transitioned flow and only partial streamlining of the model \cite{hoerner1958fluid}.
 
 Finally, the trajectory during launch (position $x(t)$, $y(t)$, $z(t)$ and rotation $R(t)$) once non-dimensionalized can only be a function of these input groups. For example for $x(t)$:
 \begin{align}
\tilde{x}(\tilde{t}) &= \frac{x(\tilde{t})}{L} = f_x(\tilde{t},\textit{Fr},\tilde{U}_\text{vehicle},\tilde{m},\tilde{d},\tilde{I})
\end{align}
Accordingly, the trajectory of the current SQUID prototype launched at 35mph from a 50mph vehicle ($\textit{Fr}=9.4$, $\tilde{d}=0.27$, $\tilde{U}_\text{vehicle}=1.4$) can be used to predict trajectories for scaled prototypes. For example, a 2x scale model (i.e. 8 times the weight, 32 times the inertia, etc.) launched at 50mph from a 70mph vehicle will match these same non-dimensional inputs. Such a model would therefore follow the same trajectory scaled by 2x the distance and take $\sqrt{2}$ times amount of time to do so.

\section{Conclusion} \label{sec:Conclusion}
The SQUID prototype has proven capable of ballistic launch, stable midair deployment, and controlled flight under manual control. A functional prototype was built and tested using commercial electronic components with a 3D printed structure. Several fully operational flights showed the benefits of the approach, both from static and mobile vehicles. 
Future work on the project will involve increasing automation of the launch process. The trigger mechanism can be activated automatically after a predefined amount of time after the flight controller registers the massive launch acceleration, rather than manually by the pilot. Additionally, while the vehicle is capable of autonomous flight using the GPS, a non-metallic launcher tube would allow GPS use from launch. The final task is adapting the SQUID concept to larger scale Earth models or mission-specific versions for Mars and Titan~
\cite{bib:titan}.


\section*{ACKNOWLEDGMENT}
The research described in this paper was performed by the California Institute of Technology and the Jet Propulsion Laboratory, California Institute of Technology, under a contract with the National Aeronautics and Space Administration with funding provided by the DARPA Mobile Force Protection Program. The authors thank Andrew Ricci, Anushri Dixit, Carl Folkestad, Joe Jordan and Reza Nemovi for their support during field testing.

\bibliographystyle{IEEEtran}
\bibliography{references}

\end{document}